\documentclass[11pt,a4paper]{article}
\usepackage{jcappub}
\usepackage{bm}
\usepackage{bbm}
\usepackage{amsmath}
\usepackage{etoolbox}
\usepackage{float}
\usepackage{mhchem}

\def\circa#1{\,\raise.3ex\hbox{$#1$\kern-.75em\lower1ex\hbox{$\sim$}}\,}

\newcommand{\beq}{\begin{equation}}
\newcommand{\eeq}{\end{equation}}

\usepackage{etoolbox}

\makeatletter
\gdef\@fpheader{}
\makeatother

\begin{document}

\makeatletter

\title{Dark Matter prospects with COSI: ALPs, PBHs and sub-GeV Dark Matter}

\author[a,b]{Andrea Caputo,}
\author[b,c,d ]{Michela Negro,}
% \author[0000-0002-6548-5622]{Michela Negro}
\author[f,g]{Marco Regis,}
\author[g]{Marco Taoso}

\affiliation[a]{School of Physics and Astronomy, Tel-Aviv University, \\ Tel-Aviv 69978, Israel}
\affiliation[b]{University of Maryland, Baltimore County, Baltimore, MD 21250, USA}
\affiliation[c]{NASA Goddard Space Flight Center, Greenbelt, MD 20771, USA}
\affiliation[d]{Center for Research and Exploration in Space Science and Technology, NASA/GSFC, Greenbelt, MD 20771, USA}

\affiliation[f]{Dipartimento di Fisica, Universit\`{a} di Torino, via P. Giuria 1, I--10125 Torino, Italy}
\affiliation[g]{Istituto Nazionale di Fisica Nucleare, Sezione di Torino, via P. Giuria 1, I--10125 Torino, Italy}

\emailAdd{andreacaputo@mail.tau.ac.il}
\emailAdd{mnegro1@umbc.edu}
\emailAdd{marco.regis@unito.it}
\emailAdd{marco.taoso@to.infn.it}

\abstract{We study the prospects in the search of dark matter offered by the newly selected NASA MeV mission COSI (Compton Spectrometer and Imager). This instrument is designed and optimized to detect spectral lines, and we show it offers an exquisite possibility to detect dark matter directly decaying or annihilating into monochromatic gamma-rays. This is the case, for example, for axion-like particles (ALPs) which undergo decay into two photons. Furthermore, we show that COSI can lead to important progress in the quest for primordial black holes (PBHs) dark matter, through measurements of the 511 keV line from the positrons produced via Hawking evaporation.
We also outline opportunities for the search of continuum signals, such as those expected from sub-GeV dark matter annihilation/decay into leptons and PBH evaporation into photons. We find that also in this case COSI can lead to improvements of current bounds.}

\maketitle

%\clearpage

\section{Introduction}

The MeV gamma-ray energy range is one of the frontiers in observational astronomy, with huge implications for the understanding of high-energy astrophysical phenomena and involving significant activity for the next-generation of instruments. Indeed, several mission concepts have been developed in the past decades with main focus to observe the so-called \textit{MeV gap}, a `gap' in sensitivity between $\sim$100 keV and $\sim$100 MeV. Some examples that appeared in literature are AMEGO ~\cite{AMEGO:2019gny}, (e-)ASTROGAM ~\cite{e-ASTROGAM:2017pxr, DeAngelis:2021esn}, COSI \citep{COSIofficial}, AMEGO-X ~\cite{Caputo:2022xpx}, ETCC~\cite{Takada:2021iug}. These mission concepts aim to improve the sensitivity in the MeV band which is currently covered only by the observations of the COMPTEL \citep{COMPTEL} on board of the NASA's Compton Gamma-ray Observatory (CGRO, \cite{CGRO}, decommissioned after 9 years of operation in 2000) and INTEGRAL/SPI (\citep{SPI}, which out-perform COMPTEL sensitivities below 2 MeV). 

In this work we focus our attention on the mission concept that have been recently selected by NASA as the next SMEX: the Compton Spectrometer and Imager, COSI \citep{COSIofficial, BeechertCOSICalib}. This mission, currently in Phase B of the NASA's Project Life Cycle, will have important implications for physics and astrophysics, from stellar nucleosynthesis to the production of antimatter (positrons) in our Galaxy, to jets formation and mechanisms in gamma-ray bursts, blazars and compact objects, and -- as we prove in this work -- it can also lead to dark matter (DM) discovery. 

COSI is a wide field of view gamma-ray telescope that allows unprecedented high-resolution spectrometry in the 0.2--5 MeV energy range thanks to cryogenic germanium detectors. COSI detects high-energy photons that Compton scatter in the detectors and, hence, it has inherited imaging capabilities with angular resolution that well meet the requirements of 3.8$^\circ$ full width at half maximum (FWHM) at 0.511 MeV and 2$^\circ$ FWHM at 1.809 MeV. Scheduled to be launched in 2027, COSI will survey the sky orbiting  Earth at 550 km altitude along an almost equatorial orbit, with a North-South rocking, which will assure a daily full-sky coverage. 

The design of the instrument is optimized to resolve the antimatter distribution in the center of our galaxy though the 511 keV $e^+e^-$ annihilation emission line; to provide major advances in nuclear line studies, in particular from $^{26}$Al, $^{60}$Fe and $^{44}$Ti radionuclides, tracing element formation, past core-collapse supernovae activity, and new supernovae explosions, respectively; to measure polarization of photons emitted in extreme environments which will provide insights on the gamma-ray production mechanisms and the emission regions geometry; to probe gravitational-waves counterparts. Obviously, COSI's potential is not limited to these main objectives, and in this paper we investigate COSI's capabilities in the search for DM.

In fact, various DM models predict the production of gamma-ray photons in the MeV range, see e.g.~\cite{Boehm:2002yz,Beacom:2004pe,Finkbeiner:2007kk,Essig:2009jx,Essig:2013goa,Boddy:2015efa}. This is the case of direct DM decay or annihilation into pairs of photons, which produces narrow lines for which COSI will have an exquisite sensitivity. Other models, such as primordial black holes (PBHs)~\cite{Carr:2020xqk,Green:2020jor,Sasaki:2018dmp,Villanueva-Domingo:2021spv, Ray:2021mxu} or DM decay/annihilation into leptons~, will instead produce a continuum photon signal, which can be investigated as well by COSI, even though with a somewhat weaker constraining power. All these scenarios can inject positrons that, after cooling, can annihilate with thermal electrons to produce a 511 keV line in the Galaxy~\cite{Knodlseder:2005yq}, and this possibility can be tightly constrained by COSI.

In this study we derive the prospects for DM searches with COSI for all the DM candidates and signatures mentioned above.

The paper is organized as follows. In Sec.~\ref{Sec:Lines} we explore COSI sensitivity for line signals coming from DM decay and annihilation. We consider various targets such as the Galactic Center (GC), the Large Magellanic Cloud (LMC) and the dwarf spheroidal galaxy Draco. Then we move to the continuum sensitivity in Sec.~\ref{Sec:Continuum}; we consider, in particular, signals from PBH Hawking radiation and final state radiation from electrons/positrons originating from DM decay and annihilation. The case of the 511 keV line is analysed in Sec.~\ref{Sec:511keV}. We finally conclude and outline future directions in Sec.~\ref{Sec:Conclusion}. 

\section{Spectral line signatures}\label{Sec:Lines}

We first study spectral line signals, for which COSI is optimized. 
%A narrow line is expected in DM models where DM either decays or annihilates directly into two photons. In the first case, the energy of the associated photons will be $E_\gamma = m_\chi/2$, where $m_\chi$ is the mass of DM, while in the second case $E_\gamma = m_\chi$. In the following we study both scenarios and consider various astrophysical targets, namely the Galactic Center (GC), the dwarf spheroidal galaxy Draco and the Large Magellanic Cloud (LMC).
In all such targets, the DM has a velocity dispersion $\sigma/c < 10^{-3}$, which sets the intrinsic width of the spectral line, while the COSI spectral resolution is $\Delta E_\gamma/E_\gamma>10^{-3}$ in the whole energy range of operation. 
Therefore, we can assume the DM signal to be contained in a single COSI energy bin.

\begin{figure}[h]
\begin{center}
  \includegraphics[width=0.49\textwidth]{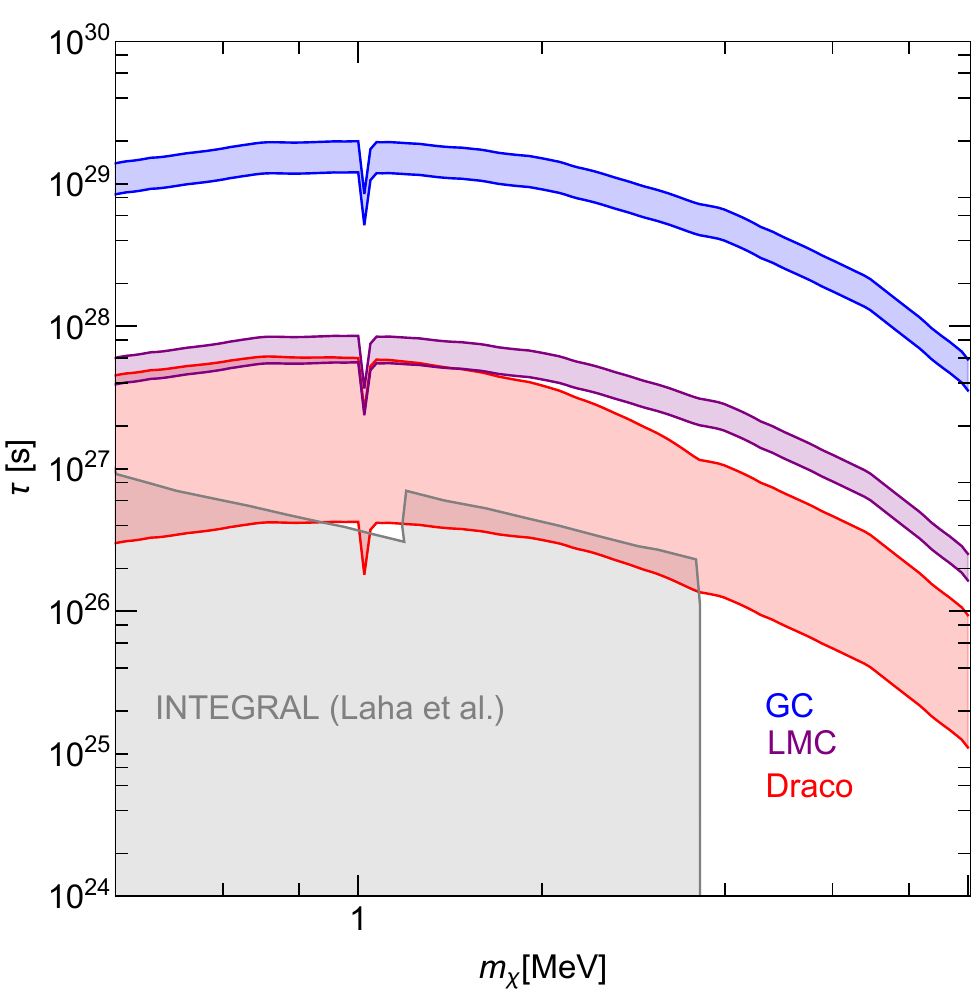}
  \includegraphics[width=0.49\textwidth]{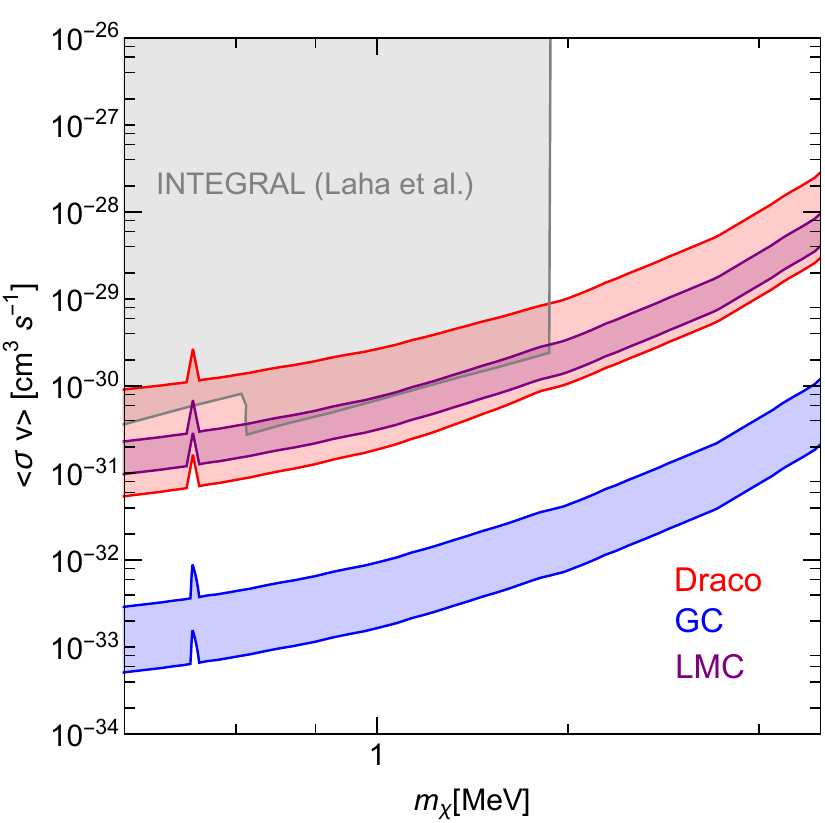}
  \vspace{-0.2cm}
  \caption{3$\sigma$ projected sensitivities (for 2 years of observation) for spectral line signatures in the three targets under consideration (GC, LMC and Draco), for decaying DM on the left and annihilating DM on the right. The shaded regions span the two DM density profiles considered in our analysis, see text for more details.
  The gray shaded region shows existing limits from \cite{Laha:2020ivk}. COSI will therefore improve over existing bounds in both cases, probing \textit{larger} lifetimes for decaying DM and \textit{smaller} cross-sections for annihilating DM. Even more stringent bounds are expected when considering a larger portion of the sky analyzed with a full template analysis similarly to what done recently in~\cite{Calore:2022pks} with INTEGRAL data.}
  \label{fig:line}
\end{center}
\end{figure}

\subsection{Decay}
\label{sec:decay}

Let us consider a particle DM with mass $m_\chi$ decaying directly into two photons with a rate $\Gamma_d$ (this may be the case of axion-like particles, ALPs).
%sterile neu \item one first example is an hidden photino DM %which decays to a photon and a gravitino;
%   \item another interesting possibility is gravitino DM. In this case one can look for the decay into a neutrino and a photon, a process induced by a mixing between the photino and the neutrino;
%    \item dipole DM, 
%    \item scalar or pseudoscalar DM decaying into two photons (this would be the case, for example, of an axion-like particle, ALP).
%Consider DM halo of a certain astrophysical object to be in the field of view of a telescope, having an angular resolution $\theta$.
We shall focus on the signal produced by the DM halo of a certain astrophysical object in the the field of view of the telescope.
%If the energy resolution of the instrument is much larger than the velocity dispersion of the DM in the target, as for the cases that we are going to consider, 
As already anticipated, the energy resolution of the instrument is much larger than the velocity dispersion of the DM in the targets that we are going to consider. Therefore,
the flux of photons received by the telescope is entirely contained in the bin at energy $E_\gamma=m_\chi/2$ and reads:
\begin{equation}
\Phi(E_\gamma) = \frac{\Gamma_d}{4 \pi m_\chi}D(\theta)\times 2\, ,
\end{equation}

where $D(\theta)$ is the so-called D-factor, which measures the amount of DM along the line of sight in the targeted object:
\begin{equation}
    D(\theta) = 2\pi\, \int_0^{\theta} d\theta' \sin(\theta') \int_{\rm l.o.s.} dl \, \rho(l, \theta)\;,
    \label{eq:dec}
\end{equation}
with $\rho$ the DM density distribution and $\theta$ the angular distance from the center of the target.
For the moment, we consider the integral over a single observational angular beam, i.e. we take $\theta$ to be the half width
at half maximum (HWHM) of the angular resolution measure (ARM) of the Compton telescope.
Notice that here we assume $\chi \rightarrow \gamma \gamma$, however in some models the decay can actually produce only one detectable photon, $\chi \rightarrow X \gamma$. Assuming $m_X\ll m_\chi$, the latter case can be described by the same Eq.~(\ref{eq:dec}) without the factor of 2. This is actually the case for a few motivated examples~\cite{Essig:2013goa}, such as dipole DM, where the dipole operator $\frac{\lambda}{\Lambda} \bar{\chi}_2 \sigma^{\mu \nu} \chi_1 F_{\mu \nu}$ mediates the decay $\chi_2 \rightarrow \chi_1 \gamma$, hidden photino DM which decays to a photon and a gravitino, gravitino DM decaying into a neutrino and a photon, sterile neutrinos DM decaying into a photon and a light active neutrino.

To derive the projected sensitivity on the DM lifetime $\tau=1/\Gamma_d$ as a function of the DM mass $m_\chi$, we consider the 3$\sigma$ line sensitivity for point sources $\Phi_{\rm exp}(E_\gamma)$ reported in Figure 2 of \citep{Tomsick:2021wed}, and corresponding to 24 months of COSI survey.
For point-like sources we can thus directly compare the outcome of Eq.~(\ref{eq:dec}) with $\Phi_{\rm exp}(E_\gamma)$ to obtain the sensitivity on $\tau$:
\begin{equation}
    \tau=1.6\times 10^{28}\,{\rm s}\,\frac{D}{10^{23}\,{\rm MeV\,cm^{-2}}}\,\frac{10^{-6}\,{\rm cm^{2}\,s}}{\Phi_{\rm exp}}\,\frac{1\,{\rm MeV}}{m_\chi}\;,
    \label{eq:taudec}
\end{equation}
where $\Phi_{\rm exp}$ is evaluated at $E_\gamma=m_\chi/2$ and the angular integration for $D$ is given by the COSI angular resolution at $E_\gamma$ as specified before. For reference, we provide the values of $D$ at $1^\circ$ for the targets under investigation in Table~\ref{tab:DJ}. To bracket the uncertainty in its determination we consider two cases, one more promising and one less promising, for each target: NFW and Isothermal  profiles for LMC (with parameters from \cite{Regis:2021glv}) and GC (taking $\rho_\odot=0.4\, {\rm GeV/cm^3}$~\cite{Karukes:2019jxv} and $r_s=4.38$ kpc (Isothermal) and $r_s=24.42$ kpc (NFW) \cite{Cirelli:2010xx}), and upper and lower 68\% C.L. limits for the D-factor of Draco derived in \cite{Bonnivard:2015xpq}.

We consider the COSI angular resolution requirements at 0.511 MeV and 1.809 MeV reported in Table 1 of \citep{Tomsick:2021wed}. Then our approximation for other energies is based on a linear interpolation between 0.511 MeV and 1.809 MeV and a linear extrapolation below 0.511 MeV. We assume a flat angular resolution above 1.809 MeV.

\begin{table}[h]
\centering
 \begin{tabular}{||c| c c ||} 
 \hline
 Target & D-factor[$1^\circ$] & J-factor[$1^\circ$]  \\ 
  & ${\rm MeV\,cm^{-2}}$ & ${\rm MeV^2\,cm^{-5}}$   \\\hline\hline
 GC & $0.55-1.7\times10^{23}$ & $0.067-4.4\times10^{27}$ \\ 
 \hline
 LMC & $1.8-2.7\times10^{22}$& $1.6-5.3\times10^{25}$  \\
 \hline
 Draco & $0.52-4.5\times10^{22}$& $0.83-8.0\times10^{25}$ \\
 \hline
\end{tabular}
\caption{D- and J-factors at $1^\circ$ for the targets under investigation. The references used in the computation are  \cite{Cirelli:2010xx} (GC) but normalised to a DM density of $0.4 \, \rm GeV/cm^3$ at the Earth location~\cite{Karukes:2019jxv}, \cite{Regis:2021glv} (LMC), and \cite{Bonnivard:2015xpq} (Draco).}
\label{tab:DJ}
\end{table}

Given the relatively poor angular resolution of COSI, Draco can be considered as a point-like source. In the case of the GC and (to a lesser extent) LMC, it would be desirable to analyze the target as an extended source.
%For the LMC, one could attempt to measure the extended emission outside the central beam, but the gain in terms of constraining power is little and the uncertainty in the LMC DM profile at large angular distances are significant. 
For extended sources, dedicated simulations by the COSI Collaboration would be ideal in order to derive the telescope sensitivity. In this work, following the procedure described in \cite{Negro:2021urm} (Appendix B), we use the scaling of the sensitivity with the power of 1/4 of the ratio between the extended source area and the beam area. Thus, for extended sources, we compute the D-factor in Eq.~\ref{eq:taudec} over the source region $\Omega_{\rm src}$ and rescale the point-source sensitivity $\Phi_{\rm exp}$ by a factor $(\Omega_{\rm src}/\Omega_{\rm res})^{0.25}$, with $\Omega_{\rm res}=\pi\,{\rm (HWHM_{ARM})^2} $.
%we rescale the sensitivity with $(D_{\rm res}/D_{\rm src})^{0.25}$ where $D_{\rm res}$ is the D-factor computed integrating over the COSI angular resolution and $D_{\rm src}$ is the D-factor computed integrating over a more extended region. 
For the GC we consider a region of radius of $10^\circ$ with an improvement of more than one order of magnitude in the bound, with respect to considering a single central beam, while for the LMC we take $5^\circ$ and the effect is more limited. 
Results about the sensitivity on the DM parameter space are shown in Fig.~\ref{fig:line} (left panel).

%we can obtain an approximate estimate by considering $N$ individual pointings across the source size with $N=\Omega_{\rm src}/\Omega_{\rm res}$ (where $\Omega_{\rm src}$ is the angular size considered for the source and $\Omega_{\rm res}$ is the angular resolution of the telescope), and treating each pointing independently with a sensitivity given by the aforementioned $\Phi_{\rm exp}(E_\gamma)$. The blue dashed line in Fig.~\ref{fig:line} shows the case for a source region of radius $10^\circ$ around the GC.

Recently, Ref.~\citep{Calore:2022pks} derived strong bounds on MeV DM using INTEGRAL/SPI observations of a large Galactic region with $|\ell|<47.5^\circ$ and $|b|<47.5^\circ$. Our simple extrapolation of the COSI sensitivity for extended sources would be too naive on such a large portion of the sky, and dedicated simulations from the COSI Collaboration are in order to obtain a fair comparison. The bounds from Ref.~\citep{Calore:2022pks} are thus not included here. On the other hand, assuming the ratio between the INTEGRAL/SPI sensitivity and the COSI one for extended regions to be the same as for the case of point-like sources, we notice that less than 2-years of COSI survey (i.e., less than 6 months of integration time) should suffice to obtain a better sensitivity than the one of INTEGRAL/SPI for 65 Ms (i.e., the integration time of the data used in \citep{Calore:2022pks}) at energies around MeV.

We should also stress that a few gamma-ray lines from nuclear transitions fall in the energy range of interest. The list includes nuclear decay chains of the isotopes 
\ce{^{56} Ni} with lines arising at 0.812, 0.847 and 1.238 MeV, \ce{^{60} Fe}, with lines at 1.173 and 1.333 MeV, \ce{^{26} Al}, with a line at 1.809 MeV, \ce{^{44} Ti} with a line at 1.157 MeV, and the 511 keV line from $e^+e^-$ annihilation. Therefore, in a few frequency channels, out of the thousands that can be explored given the energy range and resolution of COSI, the DM  signal could be contaminated by these emissions.
In these cases, a proper modeling and subtraction of these backgrounds should be pursued, similarly to the approach adopted in~\cite{Calore:2022pks,Berteaud:2022tws} with INTEGRAL/SPI observations.
Such analysis is beyond the scope of our work, but we notice that COSI will provide unprecedented sensitivity to characterize these nuclear emission lines~\cite{COSIofficial}.

\begin{figure}[H]
\begin{center}
  \includegraphics[width=0.75\textwidth]{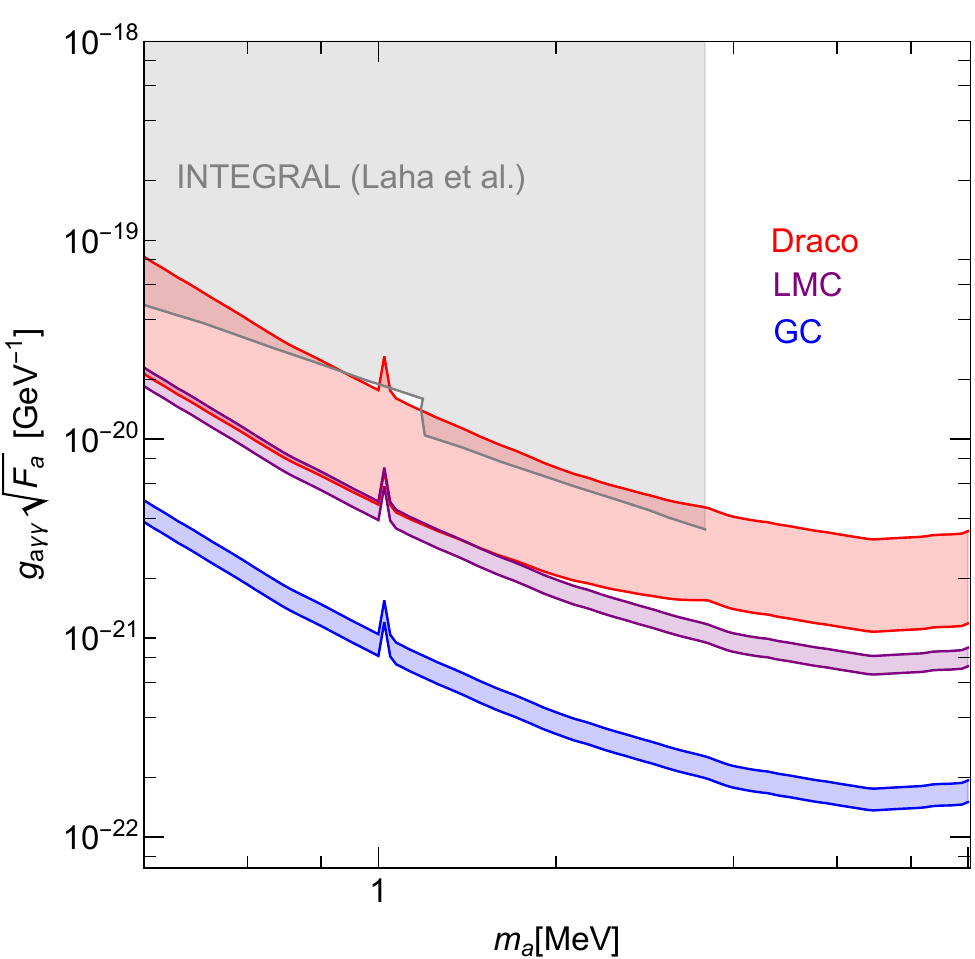}
  \vspace{-0.2cm}
  \caption{COSI prospects for ALPs constituting a fraction $F_a$ of dark matter and decaying into two photons. As in Fig.\ref{fig:line} the three different colors refer to the different targets under consideration (GC, LMC and Draco) and the shaded regions span the two DM density profiles considered in our analysis. The gray shaded region shows existing limits from~\cite{Laha:2020ivk}.}
  \label{fig:ALPsBounds}
\end{center}
\end{figure}

Finally, our bounds apply to any model in which DM radiatively decays into photons. It is however interesting to convert COSI prospects on DM lifetime in forecasts for the specific case of ALPs, with mass $m_a$ and coupling to photons $g_{a\gamma \gamma}$. In Fig.~\ref{fig:ALPsBounds} we show COSI prospects for this model assuming the ALPs constitute a fraction $F_a$ of the DM, i.e., being agnostic about the ALPs relic abundance. These limits, for the mass range of interest, apply for coupling as large as $g_{a\gamma\gamma} \simeq 10^{-10}-10^{-11} \rm GeV^{-1}$, above which the DM lifetime becomes shorter than the age of the universe. Notice that one could use Fig.~\ref{fig:ALPsBounds} to constrain any irreducible axion component that is produced in the early Universe and with a relic abundance today (as done in Ref.~\cite{Langhoff:2022bij} considering minimal freeze-in production of axions). Finally, we also highlight that axions in this mass range has been recently shown to be theoretically motivated, as they naturally arise from axiverse theories with dark non-abelian gauge groups~\cite{Foster:2022ajl}.

\subsection{Annihilation}
In this section we focus on direct DM annihilation into two photons. 
%Again, we consider the flux in a single energy bin since, in the targets under consideration, DM has a negligible dispersion velocity compared with the experimental spectral resolution.
The flux, produced at $E_\gamma=m_\chi$, reads:
\begin{equation}
\Phi(E_\gamma) = \frac{\langle \sigma_a v\rangle}{4 \pi m_\chi^2}J(\theta)\, ,
\end{equation}
where $\langle \sigma_a v\rangle$ is the velocity-averaged annihilation cross-section and the so-called J-factor is:
\begin{equation}
    J(\theta) = 2\pi\, \int_0^{\theta} d\theta' \sin(\theta') \int dl \, \rho(l, \theta)^2\;.
\end{equation}

Using the same arguments as for the decaying DM case, we arrive at the projected sensitivity on the annihilation cross-section:
\begin{equation}
    \langle \sigma_a v\rangle=1.3\times 10^{-32}\,{\rm cm^3/s}\,\frac{10^{27}\,{\rm GeV^2\,cm^{-5}}}{J}\,\frac{\Phi_{\rm exp}}{10^{-6}\,{\rm cm^{2}\,s}}\,\left(\frac{m_\chi}{1\,{\rm MeV}}\right)^2
    \label{eq:sigmaann}
\end{equation}
where $\Phi_{\rm exp}$ is evaluated at $E_\gamma=m_\chi$ and the angular integration for $J$ is given by the COSI angular resolution at $E_\gamma$ (see some reference values in Table~\ref{tab:DJ}).
Results are in the right panel of Fig.~\ref{fig:line}, for the same targets and regions of observations considered in Section~\ref{sec:decay}.

%\MR{TO BE CHECKED For annihilating DM, the emission profile is much more peaked towards the center, and the point-like source approximation essentially holds for all the three targets.}

Before moving to continuum signals, we notice that Ref.~\cite{Aramaki:2022zpw} also provided an estimate of the reach of COSI for $\langle \sigma_a v\rangle$. The values obtained by the authors (see their discussion at pag.~14 of Ref.~\cite{Aramaki:2022zpw}) are significantly more constraining than ours. We checked that this difference is due to a different choice of the J-factor, which was more optimistic in Ref.~\cite{Aramaki:2022zpw}.

\section{Continuum signals}\label{Sec:Continuum}

We now move to DM scenarios for which a continuum photon signal is expected. 
This is the case of light DM decaying or annihilating into leptons or mesons, which in turn emit radiation. 
Another possibility are PBHs, which evaporate producing an (almost) thermal spectrum of photons.
%This is the case, for example, of PBHs
%for which an (almost) thermal spectrum of photons is originated.
%expected to originate. 
%This is also the case of models in which light DM decays or annihilates into leptons or mesons, which then emit radiation. 
%For example, if DM is a ``light" scalar a decay into leptons can naturally arise. 

As we mentioned in the introduction, COSI is optimized
%in sensitivity 
for the detection of MeV lines. The continuum sensitivity suffers from a larger instrumental background, which limits the detection efficiency. The possibility to constrain the DM scenarios that we have just described are somewhat less promising than for the cases of Section~\ref{Sec:Lines}, but anyway relevant, and improving on current bounds for some masses, as we describe below.

\begin{figure}[H]
\begin{center}
  \includegraphics[width=0.55\textwidth]{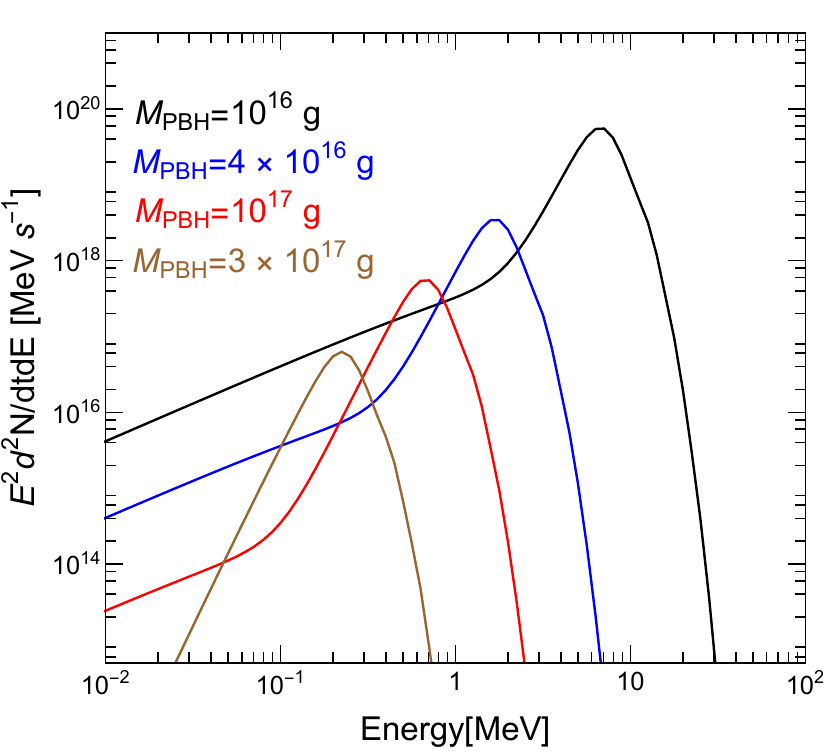}
  \vspace{-0.2cm}
  \caption{Differential photon emission rate due to Hawking radiation~\cite{Hawking:1974rv} for PBH with different masses. These spectra include both primary and secondary photon emissions (originating from other produced particles, such as electrons) as obtained using the public code BlackHawk~\cite{Arbey:2019mbc}. The smaller the PBH mass, the larger the photons peak energy, as it follows from the BH temperature~\cite{Hawking:1974rv} $ T_{BH} = \frac{1}{8 \pi G M_{BH}} \simeq \, \rm MeV \Big(\frac{10^{16} \rm g}{M_{BH}}\Big)$.
  }
  \label{fig:PBHSpectrum}
\end{center}
\end{figure}

\subsection{Primordial black holes evaporation}\label{Sec:PBHs}

The recent detection of gravitational waves from black holes mergers prompted a renewed interest in the possibility that all or a sizeable fraction of the DM in our Universe is constituted by PBHs, that is to say black holes formed in the early universe and not from stellar evolution, see e.g.~\cite{Carr:2020xqk,Green:2020jor,Sasaki:2018dmp,Villanueva-Domingo:2021spv} for recent reviews.

In his seminal work Hawking demonstrated that a black hole (BH) is not really black, but actually emits an (almost) thermal spectrum of particles with temperature~\cite{Hawking:1974rv}
\begin{equation}
    T_{BH} = \frac{1}{8 \pi G M_{BH}} \simeq \, \rm MeV \Big(\frac{10^{16} \rm g}{M_{BH}}\Big),
\end{equation}
where $G$ is the Newton gravitational constant and $M_{BH}$ the BH mass. In particular, the emission rate of a particle state of spin s and energy $\omega$ from  a Schwarzschild BH is given by
\begin{equation}
 \frac{dN_i}{d\omega dt} = \frac{g_i}{2\pi}\frac{\Gamma_i(\omega, T_{BH})}{e^{\omega/T_{BH}}-(-1)^{2s}},
\end{equation}
where $g_i$ is the number of internal dof of the i-th particle, 
%$\omega = \sqrt{p^2 + m_i^2}$ 
$\omega$ is its energy, and the function $\Gamma_i$ denotes the so-called grey factor. The latter is the dimensionless absorption probability for
the emitted species. It is in general a complicated function of $\omega$, the PBH mass, the particle's internal dof and its rest mass. In order to compute the associated photon spectrum we make use of the public code BlackHawk~\cite{Arbey:2019mbc}, including also the photons originating from unstable particles produced during the PBH evaporation (calculated using Hazma~\cite{Coogan:2020tuf} to compute the photon spectrum from decays). Denoting the
%complete 
differential photon emission rate
%flux per unit time 
by $\frac{d^2 N_\gamma}{dE_\gamma dt}$, which we show in Fig.~\ref{fig:PBHSpectrum} for a few PBH masses, 
the flux of photons with energy $E_\gamma$ received by the telescope is:
%the flux of photons received at the telescope at energy bin $E_\gamma$ will be
\begin{equation}
    \Phi(E_\gamma) = \frac{f_{\rm PBH}D(\theta)}{4\pi \, M_{\rm PBH}}\frac{d^2 N_\gamma}{dE_\gamma dt},
\end{equation}

where $M_{PBH}$ is the PBH mass , $f_{\rm PBH}$ is the fraction of DM in form of PBHs
%PBH constituting DM 
and $D(\theta)$ is the D-factor integrated over a certain source area of angular size $\theta$ as already described above. Notice that in this work we only consider a monochromatic mass distribution for the PBHs. 
Nevertheless, our analysis can be easily be adapted to a different mass function, see e.g.~\cite{Carr:2017jsz}.

For a given PBH mass we can then compare the theoretical prediction with the COSI sensitivity and estimate the value of $f_{\rm PBH}$ which can be tested. For the COSI sensitivity in the continuum we use the five points in Figure 4 of Ref.~\cite{Negro:2021urm} rescaled with a power of 1/4 of the ratio between the source area and an area of 20-degree radius, to take into account the fact that Figure 4 of Ref.~\cite{Negro:2021urm} refers to the performance for a disk-like source with 20-degree radius (see their Appendix B).

Notice also that those prospects are related to the Fermi Bubbles, i.e., to the central part of the Galaxy. Nevertheless we expect them to be similar in other directions of the sky since the Galactic foreground plays a minor role in the determination of the sensitivity, which is dominated by the instrumental background. 
The bound on the PBH fraction is therefore computed as 
\begin{equation}
\label{eq:fBH}
    f_{\rm PBH} < \Big[\frac{4 \pi M_{\rm PBH}}{D(\theta)} \times \frac{1}{d^2 N_\gamma/dE_\gamma dt} \times \Phi_{\rm exp}(E_\gamma) \times \left(\frac{\pi\,\theta^2}{\pi\,20^{\circ\,2}}\right)^{1/4}\Big]_{\rm min},
\end{equation}

where $\Phi_{\rm exp}(E_\gamma)$ is the COSI sensitivity
of Ref.~\cite{Negro:2021urm}. The subscript ``min'' means that we evaluate the above expression at the five energy bins of Figure 4 in Ref.~\cite{Negro:2021urm} and then take the minimum value. 

\begin{figure}[H]
\begin{center}
  \includegraphics[width=0.5\textwidth]{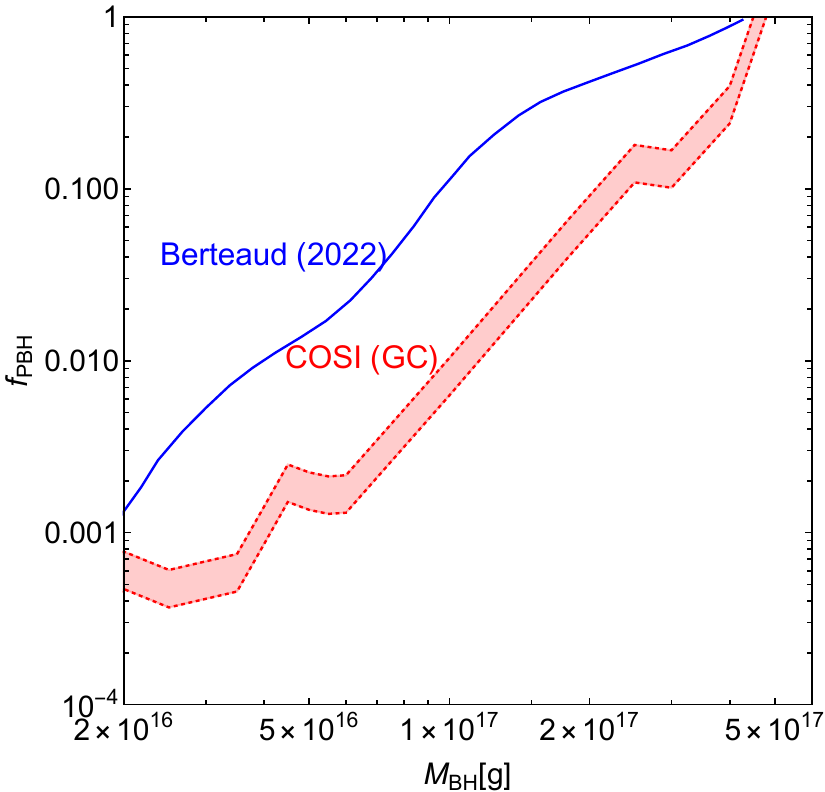}
  \vspace{-0.2cm}
  \caption{COSI prospects for the detection of PBH Hawking radiation into photons from the GC (shaded red region, which spans the two DM density profiles considered in our analysis). The solid blue line indicates the existing bounds from INTEGRAL/SPI observations, taken from Ref.~\cite{Berteaud:2022tws}.
  }
  \label{fig:PBHCosi}
\end{center}
\end{figure}

We present these prospects for the GC
%of our analysis 
as a shaded red band in Fig.~\ref{fig:PBHCosi} (spanning the two considered density profiles, as done for spectral line signals). 
For comparison we also show current bounds (solid blue line) from INTEGRAL/SPI observations as taken from Ref.~\cite{Berteaud:2022tws}. As evident, COSI will be able to improve present constraints also in this case.

\subsection{Annihilation and decay of sub-GeV DM into leptons}

%For many years DM searches have been dominated by the paradigm of
%Weakly Interacting Massive Particles (WIMPs), with masses around the GeV scale or above. However, the so-far absence of signals have questioned past paradigms and the attention turned also to lighter masses, below the GeV scale. 
%In the previous sections we have considered direct DM decays and annihilations 
%channels for which COSI will be particularly relevant, being designed to look for lines.
In Sec.~\ref{Sec:Lines}
we have considered direct DM decays and annihilations 
into photons, a scenario particularly promising for COSI, since this telescope is designed to look for gamma-ray lines. 
Nevertheless, in many motivated models DM annihilates into leptons or mesons, which in turn produce a continuum spectrum of photons. Here, for concreteness, we focus on the simple case of DM annihilating or decaying into electrons and positrons, $\chi \chi \rightarrow e^+ e^-$ or $\chi \rightarrow e^+ e^-$. For annihilations,
%the case of the annihilation 
the differential photon spectrum~\cite{Cirelli:2020bpc, Bystritskiy:2005ib,Essig:2013goa} is:
\beq
\label{eq:FSRll}
\frac{dN_{{\rm FSR} \gamma}^{e^+e^-}}{dE_\gamma} = \frac{\alpha}{\pi\beta(3-\beta^2)\,m_{\chi}} 
\left[{\cal{A}} \, \ln \frac{1+R(\nu)}{1-R(\nu)}-2\, {\cal{B}}\,  R(\nu)
\right] \,,
\eeq
where
\beq
{\cal{A}} = \left[\frac{\left(1+\beta^{2}\right)\left(3-\beta^{2}\right)}{\nu}-2\left(3-\beta^{2}\right)+2 \nu\right] \,,
\eeq
\beq
{\cal{B}} = \left[\frac{3-\beta^{2}}{\nu}(1-\nu)+\nu\right] \,,
\eeq
and: $\nu = E_\gamma/m_{\chi}$,  $\beta^2 = 1-4\mu^2$ with $\mu=m_e/(2\,m_{\chi})$, $R(\nu) = \sqrt{1-4\mu^2/(1-\nu)}$ and $m_e$ the electron mass. Given the differential photon spectrum, one can write down the expected flux of photons from a given astrophysical target (say the GC) for self-conjugated DM particles (e.g. Majorana fermions) as:
\beq
\Phi^{\rm ann}(E_\gamma) = \frac{J(\theta)}{4\pi \, m_\chi^2}\times \frac{\langle \sigma v \rangle}{2}\times \frac{dN_{{\rm FSR} \gamma}^{e^+e^-}}{dE_\gamma},
\eeq

where $J(\theta)$ is the J-factor for the observed target, $m_\chi$ the mass of the annihilating DM and $\langle \sigma v \rangle$ the annihilation cross-section. Notice that if DM is not constituted by self-conjugated particles, an additional factor $1/2$ appears.

%The decaying case is analogous. 
For decaying DM, the differential photon spectrum is the same as in Eq.~(\ref{eq:FSRll}) with the substitution $m_{\chi} \rightarrow m_{\chi}/2$ everywhere, %(in perfect agreement then with Eq.4 of Ref.~\cite{Essig:2009jx} for $m_\chi \gg m_e$) \MR{I would remove the sentence in parenthesis},
and the flux of photons reads
\beq
\Phi^{\rm dec}(E_\gamma) = \frac{D(\theta)/\tau}{4\pi \, m_\chi}\times \frac{dN_{{\rm FSR} \gamma}^{e^+e^-}}{dE_\gamma},
\eeq

where $D(\theta)$ is the D-factor and $\tau$ the DM lifetime for the decay into electron/positron pairs.

\begin{figure}[H]
\begin{center}
  \includegraphics[width=0.45\textwidth]{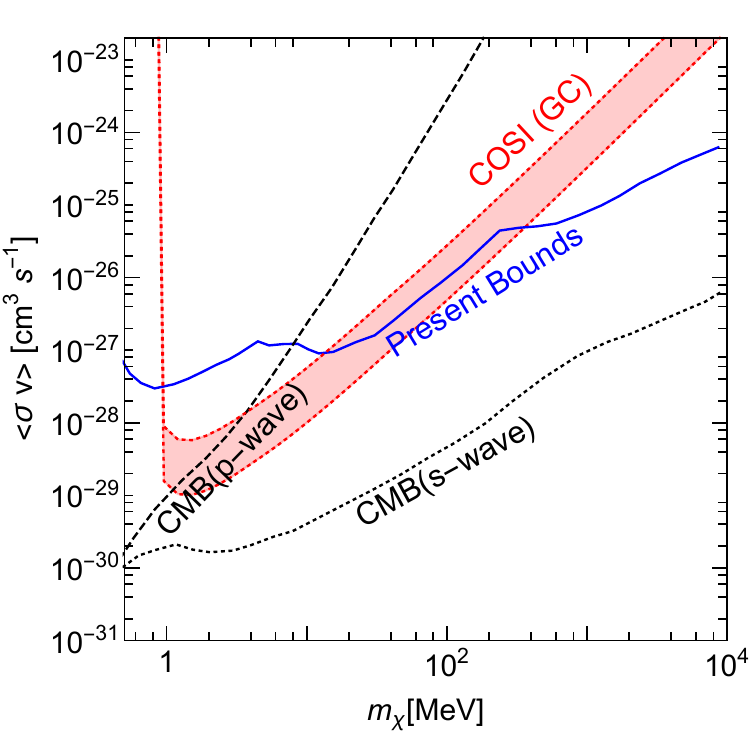}
  \includegraphics[width=0.44\textwidth]{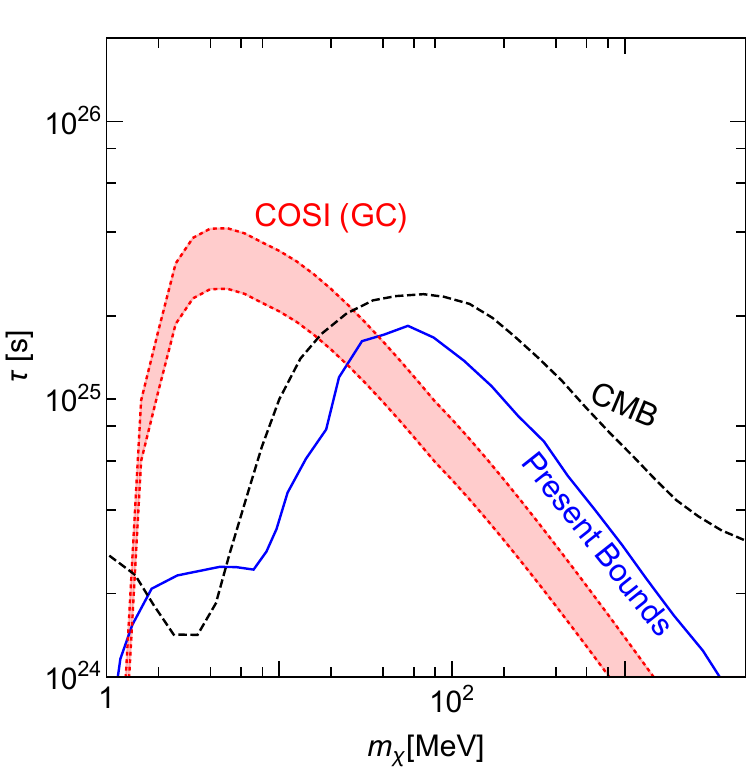}
  \vspace{-0.2cm}
  \caption{(\textbf{Left}). Projected sensitivity on the DM velocity-averaged annihilation cross-section $\langle \sigma v \rangle$ for the process $\chi \chi \rightarrow e^+ e^-$. The COSI prospects for the GC are shown as a shaded band (spanning the two considered density profiles, as in Fig.~\ref{fig:line}), current gamma-ray bounds are in solid blue (taken from Figure 1 of Ref.~\cite{Coogan:2021rez}) and CMB bounds for s-wave (dotted) and p-wave (dashed) annihilation are in black. (\textbf{Right}) Sensitivity on the lifetime of DM decaying into $e^+ e^-$. The color code is as in the left panel.
  %COSI prospects are again in red, existing gamma-ray bounds in blue and CMB limits in black.
  }
  \label{fig:EEbound}
\end{center}
\end{figure}

We then proceed to derive the COSI projected sensitivities as for the PBH case. For 2 years of mission time we derive the prospects showed in Fig.~\ref{fig:EEbound}. COSI forecasts for the GC are shown again as a shaded red band, while present bounds from other experiments such as COMPTEL, EGRET, Fermi and INTEGRAL are shown in solid blue (taken from Figure 1 of Ref.~\cite{Coogan:2021rez}). For comparison we also show the constraints from CMB temperature and polarization anisotropies as black lines~\cite{Coogan:2021rez, Slatyer:2016qyl}.

\section{COSI, dark matter and the Galactic Center 511 keV gamma-ray line}\label{Sec:511keV}

The presence of a 511 keV line from the GC is a long-standing issue, with its first detection dating back to the 1970's \cite{1972ApJ...172L...1J,1978ApJ...225L..11L}.
More recently, the INTEGRAL/SPI experiment observed an emission compatible with the annihilation of $\sim (1.5 \pm 0.1) \cdot 10^{43}$ positrons/s~\cite{Knodlseder:2005yq} in a region within $\sim 1$ kpc of the GC. Among conventional sources, there are radioactive ejecta from stars, supernovae and gamma-ray bursts,  positrons from pair creation near pulsars or from p-p collisions associated with cosmic
rays~\cite{Prantzos:2010wi}. Most of the solutions, however, are not completely satisfactory to fully explain the excess. This motivated the investigation of more ``exotic'' alternatives, some of which trace back to DM to provide the required large number of positrons. In fact, as discussed in the previous two sections, there are many DM models for which the production of $e^+e^-$ is expected. 
In these scenarios, low-energy positrons would then annihilate with ambient electrons producing the 511 keV gamma-ray line. 
%These positrons would then annihilate with ambient electrons and -- if low energetic enough -- produce the characteristic observed gamma-ray line. 

Given its exquisite spectral resolution,
%precision for spectral analysis, 
COSI will be instrumental in characterizing the 511 keV line at the GC and therefore constraining DM signals.
%help to characterize the 511 keV line at the GC and therefore to constrain DM signals.
%place better constraints on DM models. 
Let us start considering PBH DM as in Sec.~\ref{Sec:PBHs}. Given the ``democratic'' nature of gravity, a PBH 
%of a given mass will 
generates all possible spectra of particles energetically allowed, and therefore for small enough PBH masses, we also expect a copious production of $e^+e^-$, which would then contribute to the 511 keV line~\cite{Laha:2019ssq, DeRocco:2019fjq, Fuller:2017uyd}. In order to match the excess, the flux of photons should be $\Phi_{\rm 511} \sim 10^{-3} \rm ph/cm^2/s$, which in turn corresponds to a positron luminosity of $L_{e^+} \sim 2\times 10^{43} \, \rm e^+/s$~\cite{Knodlseder:2005yq}. However, the injected positrons cannot be too energetic. In fact, a sizeable emission of positrons with energies above $\sim 3$ MeV could produce an excess of high energy photons due to inflight annihilation~\cite{Beacom:2005qv}.
Nevertheless, for the PBH masses that we are considering, we checked that energetic positrons ($> 3$ MeV) only accounts for a small fraction of the total luminosity and therefore this constraint does not apply.

As for the photon spectrum in Sec.~\ref{Sec:PBHs}, we make use of the public code BlackHawk~\cite{Arbey:2019mbc} to generate the positron differential rate $dN_{e^+}/d\omega dt$, including both the primary and the secondary contributions. Therefore, the number of positrons from PBH evaporation per unit of time in the energy range of interest can be computed as
\begin{equation}\label{Eq:PositronPBH}
    L_{e^+}^{\rm PBH} = \frac{f_{\rm PBH}}{M_{\rm PBH}} \int dV \rho(r) \int_{\rm m_e}^{\rm \infty} d\omega \frac{dN_{e^+}}{d\omega dt},
\end{equation}
where for the DM density distribution $\rho$ we have adopted a NFW profile normalized as in Sec.~\ref{Sec:Lines}.
Concerning the volume integral, we assume, as done in previous analyses~\cite{Laha:2019ssq}, that all the positrons emitted within 1.5 kpc from the GC contribute to the 511 keV gamma-ray emission, i.e., the integral is limited to distances $<1.5$ kpc.
Then, Eq.~(\ref{Eq:PositronPBH}) can be equated to the experimental measurement $L_{e^+} \sim 2\times 10^{43} \, \rm e^+/s$~\cite{Knodlseder:2005yq} to obtain an upper limit on $f_{\rm PBH}$ for a given PBH mass.
%for a the chosen DM profile. 
We notice that the morphology of the signal is not compatible with DM decay (or BH evaporation in this case)~\cite{Vincent:2012an,Ascasibar:2005rw}, but one can still use the measurements to draw limits. A full morphological analysis is likely to provide more stringent bounds. 

\begin{figure}[H]
\begin{center}
  \includegraphics[width=0.55\textwidth]{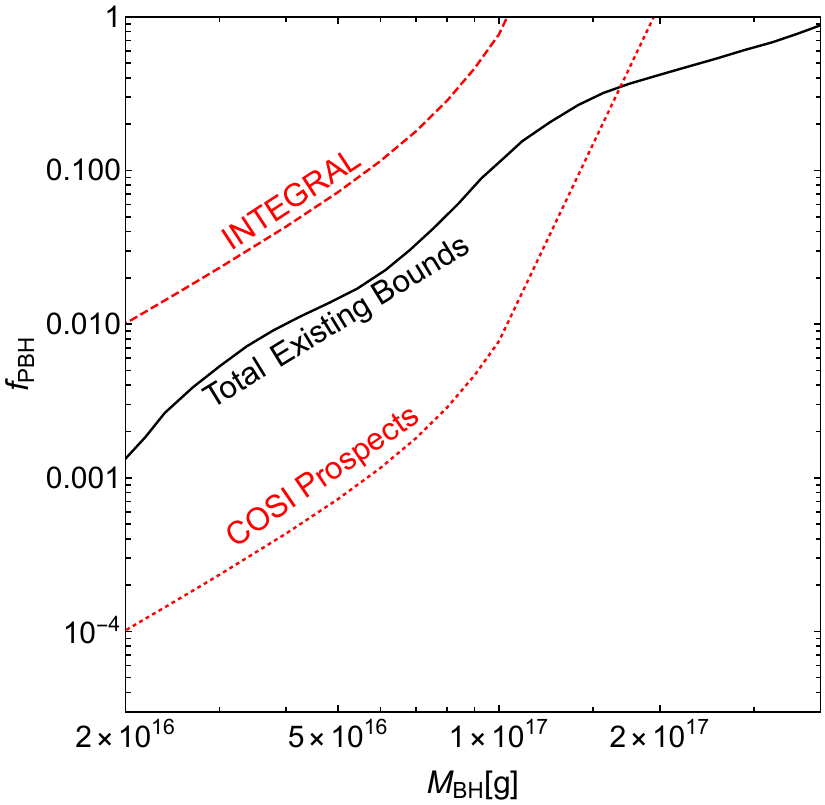}
  \vspace{-0.2cm}
  \caption{Limit on the fraction of PBH dark matter from the 511 keV emission at the GC. The dashed red curve is obtained following the recipe of Ref.~\cite{Laha:2019ssq}, but with a full relativistic treatment for the PBH evaporation spectrum.
  The solid black line indicates current bounds on $f_{\rm PBH}$ from other observations~\cite{Berteaud:2022tws}, while the dotted red curve is the prospect for the COSI mission.
  }
  \label{fig:PBH511}
\end{center}
\end{figure}

Following the outlined procedure we obtained the projected limits shown with a dashed red line in Fig.~\ref{fig:PBH511}. The results are reasonably similar to those of Ref.~\cite{Laha:2019ssq} and the (small) differences are likely due to the fact that Ref.~\cite{Laha:2019ssq} used non-relativistic approximated form for the spectra of fermions produced from PBH evaporation.

We now explore the potential of COSI to constrain a DM contribution to the 511 keV signal, under the assumption that this emission is produced by astrophysical sources, whose nature and properties will be pinpointed by COSI. Therefore, in order to constrain PBHs evaporation, we compare the corresponding signal with the sensitivity of COSI at 511 keV, i.e. the accuracy up to which the astrophysical explanation can be tested.
For a region within $5^\circ$ from the GC, i.e. the size of the detected signal, the $3\sigma$ COSI line sensitivity is $\Phi_{\rm 511}^{\rm COSI} \sim 1 \times 10^{-5} \rm ph/cm^2/s$, obtained again from Ref.~\cite{Tomsick:2021wed}, and re-scaling the point-source sensitivity as described above. This corresponds to a positron luminosity $L_{e^+} \sim 2\times 10^{41} \, \rm e^+/s$.
Using this information, we obtain the projected limits shown with a dotted red curve in Fig.~\ref{fig:PBH511}, which are roughly 2 orders of magnitude more constraining than current bounds. 
Clearly, the simple estimate presented here is affected by several sources of uncertainty. First, the propagation of low energy positrons at the GC is not well known. Then, our assessment of the constraining power of COSI might be optimistic (i.e., data perfectly fitted by an astrophysical explanation), and a precise determination of maximum fraction of the 511 keV emission which can be ascribed to DM signals will be possible only with COSI data at hands. The actual bound will be somewhere in between the dashed and the dotted curves in Fig.~\ref{fig:PBH511}.
Our calculations are in any case an indication that COSI will have the potential to strongly challenge PBH DM in the $10^{16}-10^{17}$ g mass range.   

Finally, the same procedure adopted for PBH evaporation can also be applied to DM annihilations or decays into electrons and positrons. In these two cases the positron fluxes read
\begin{eqnarray}
    \Phi_{e^+}^{\rm ann} = \frac{1}{2} \frac{4\pi}{m_\chi^2} \langle \sigma v\rangle \int_0^{\rm 1.5 kpc} dr \, r^2\rho_{\rm DM}^2(r),\\
    \Phi_{e^+}^{\rm dec} = \frac{4\pi}{m_\chi} \frac{1}{\tau} \int_0^{\rm 1.5 kpc} dr \, r^2\rho_{\rm DM}(r).
\end{eqnarray}
For small enough DM masses, below a few MeV, the positrons will not escape the GC and annihilating with ambient electrons will
%finally all 
contribute to the 511 keV line signal. Then, one can impose $\Phi_{e^+}^{\rm ann, dec} < 2\times 10^{43} \rm \,e^+/ {\rm s}$ ($ 2\times 10^{41} \, \rm e^+/ \,{\rm s}$) obtaining the bounds
%(simplistic, but likely conservative) bounds
\begin{eqnarray}
    \langle \sigma v\rangle &\lesssim& 2.3\times 10^{-30} \, {\rm cm}^3 {\rm s}^{-1} \Big(\frac{m_\chi}{\rm MeV}\Big)^2,\; \;\; \Big[2.2\times 10^{-32} \, {\rm cm}^3 {\rm s}^{-1} \Big(\frac{m_\chi}{\rm MeV}\Big)^2\Big]\nonumber\\
    \tau &>&1.1\times 10^{26} \, {\rm s} \, \Big(\frac{\rm MeV}{m_\chi}\Big),\;\;\; \Big[1.1\times 10^{28} \, {\rm s} \, \Big(\frac{\rm MeV}{m_\chi}\Big)\Big].
    \label{eq:DM511}
\end{eqnarray}

These limits, derived with a very simple procedure, are overall very similar to those obtained in Ref.~\cite{Vincent:2012an} where the authors attempted a more sophisticated analysis, modelling possible astrophysical sources contributing to the 511 keV emission. In Eq.~(\ref{eq:DM511}), in parenthesis we also report the projected limits corresponding to the COSI sensitivity, similarly to what done in Fig.~\ref{fig:PBH511} with the dotted line. We do not show the bounds of Eq.~(\ref{eq:DM511}) in Fig.~\ref{fig:EEbound}, as it is difficult to precisely draw the mass range over which they apply robustly. For masses below a few MeV they can be confidently taken and result to be the strongest ones, competing with CMB constraints even for s-wave annihilations. On the other hand, for larger masses, the description of cooling and diffusion of energetic positrons in the Galaxy becomes crucial, adding a model dependency to the bound.

\section{Conclusions}\label{Sec:Conclusion}
% \MR{To be extended/improved: please go ahead}

Scheduled to be launched in 2027, COSI will provide us with spectroscopic images of the sky at MeV energy with unprecedented sensitivity.
In this work, we discussed how to use these data to constrain different types of DM candidates.
We showed that COSI will offer an exquisite possibility to detect DM 
directly decaying or annihilating into monochromatic gamma-rays, as for the case of ALP DM.
%in the form, e.g., of ALPs, directly decaying or annihilating into gamma rays. 
Fig.~\ref{fig:line} shows that the projected sensitivities for spectral line signatures can lead to a significant improvement of the current bounds.
% on the decay and annihilation rates.

We also computed the expected sensitivity for continuum signals, such as those produced by PBH evaporation or sub-GeV DM annihilation/decay into leptons. 
For those scenarios, we found again that the projected limits are more stringent than existing limits.

Finally, we discussed how 
%to use 
measurements of the 511 keV line emission at the GC can constrain the injection of positrons from DM.
%by DM models. 
Through this observable, COSI will have the potential to significanly improve current limits, and challenge the hypothesis of PBH DM to extremely small fractional abundances in the $10^{16}-10^{17}$ g mass range.
%be able to  challenge the hypothesis of PBH DM to extremely small fractions in the $10^{16}-10^{17}$ g mass range.

The work is intended to provide a first assessment of COSI capabilities related to DM searches. More details about the final design and simulated instrumental responses will be released in the next years by the COSI Collaboration, prior to the stellite launch, and this will allow an improved determination of the projected sensitivities.

%%%%%%%%%%%%%%%%%%%%%%%%%%%%%%%%%
\section*{Acknowledgements}
%%%%%%%%%%%%%%%%%%%%%%%%%%%%%%%%%
We want to thank the COSI dark matter topical working group, and in particular we thank Shigeki Matsumoto, Tom Melia, Thomas Siegert, Tadayuki Takahashi for the fruitful discussions and kind support; John Tomsick and Andreas Zoglouer for crucial insights about the COSI instrument and capabilities. AC also thanks William De Rocco for useful discussions about the 511 keV lines and PBH dark matter. AC is supported by the Foreign Postdoctoral Fellowship Program of the Israel Academy of Sciences and Humanities and also acknowledges support from the Israel Science Foundation (Grant 1302/19), the US-Israeli BSF (Grant 2018236), the German-Israeli GIF (Grant I-2524-303.7) and the European Research Council (ERC) under the EU Horizon 2020 Programme (ERC-CoG-2015-Proposal n. 682676 LDMThExp). 
MN acknowledges NASA support under award number 80GSFC21M0002.
MR acknowledges support by the PRIN research grant ``From  Darklight  to  Dark  Matter: understanding the galaxy/matter connection to measure the Universe'' No. 20179P3PKJ funded by MIUR, and by the research grant TAsP (Theoretical Astroparticle Physics) funded by Istituto Nazionale di Fisica Nucleare (INFN).
MT acknowledges the research grant ``The Dark Universe: A Synergic Multimessenger Approach No. 2017X7X85'' funded by MIUR, and the project ``Theoretical Astroparticle Physics (TAsP)'' funded by Istituto Nazionale di Fisica Nucleare (INFN).

\bibliographystyle{bibi}
\bibliography{biblio.bib}

\providecommand{\href}[2]{#2}\begingroup\raggedright\begin{thebibliography}{10}

\bibitem{AMEGO:2019gny}
{\scshape AMEGO} Collaboration, R.~Caputo et~al., \emph{{All-sky Medium Energy
  Gamma-ray Observatory: Exploring the Extreme Multimessenger Universe}},
  \href{https://arxiv.org/abs/1907.07558}{{\ttfamily 1907.07558}}.

\bibitem{e-ASTROGAM:2017pxr}
{\scshape e-ASTROGAM} Collaboration, M.~Tavani et~al., \emph{{Science with
  e-ASTROGAM: A space mission for MeV\textendash{}GeV gamma-ray astrophysics}},
  \href{https://doi.org/10.1016/j.jheap.2018.07.001}{\emph{JHEAp} {\bfseries
  19} (2018) 1} [\href{https://arxiv.org/abs/1711.01265}{{\ttfamily
  1711.01265}}].

\bibitem{DeAngelis:2021esn}
A.~De~Angelis et~al., \emph{{Gamma-ray astrophysics in the MeV range: The
  ASTROGAM concept and beyond}},
  \href{https://doi.org/10.1007/s10686-021-09706-y}{\emph{Exper. Astron.}
  {\bfseries 51} (2021) 1225}
  [\href{https://arxiv.org/abs/2102.02460}{{\ttfamily 2102.02460}}].

\bibitem{COSIofficial}
J.~{Tomsick}, A.~{Zoglauer}, C.~{Sleator}, H.~{Lazar}, J.~{Beechert},
  S.~{Boggs}, J.~{Roberts}, T.~{Siegert}, A.~{Lowell}, E.~{Wulf}, E.~{Grove},
  B.~{Phlips}, T.~{Brandt}, A.~{Smale}, C.~{Kierans}, E.~{Burns},
  D.~{Hartmann}, M.~{Leising}, M.~{Ajello}, C.~{Fryer}, M.~{Amman}, H.-K.
  {Chang}, P.~{Jean} and P.~{von Ballmoos}, \emph{{The Compton Spectrometer and
  Imager}},  in \emph{Bulletin of the American Astronomical Society}, vol.~51,
  p.~98, Sept., 2019, \href{https://arxiv.org/abs/1908.04334}{{\ttfamily
  1908.04334}}.

\bibitem{Caputo:2022xpx}
R.~Caputo et~al., \emph{{The All-sky Medium Energy Gamma-ray Observatory
  eXplorer (AMEGO-X) Mission Concept}},
  \href{https://arxiv.org/abs/2208.04990}{{\ttfamily 2208.04990}}.

\bibitem{Takada:2021iug}
A.~Takada et~al., \emph{{First Observation of the MeV Gamma-Ray Universe with
  Bijective Imaging Spectroscopy Using the Electron-tracking Compton Telescope
  on Board SMILE-2+}},
  \href{https://doi.org/10.3847/1538-4357/ac6103}{\emph{Astrophys. J.}
  {\bfseries 930} (2022) 6} [\href{https://arxiv.org/abs/2107.00180}{{\ttfamily
  2107.00180}}].

\bibitem{COMPTEL}
V.~{Schoenfelder}, H.~{Aarts} and K.~Bennett, \emph{{Instrument Description and
  Performance of the Imaging Gamma-Ray Telescope COMPTEL aboard the Compton
  Gamma-Ray Observatory}}, \href{https://doi.org/10.1086/191794}{\emph{{ApJS}}
  {\bfseries 86} (1993) 657}.

\bibitem{CGRO}
{Compton Gamma-Ray Observatory}.
  \url{https://heasarc.gsfc.nasa.gov/docs/cgro/cgro.html}, 1991.

\bibitem{SPI}
G.~{Vedrenne}, J.~P. {Roques}, V.~{Sch{\"o}nfelder}, P.~{Mandrou}, G.~G.
  {Lichti}, A.~{von Kienlin}, B.~{Cordier}, S.~{Schanne}, J.~{Kn{\"o}dlseder},
  G.~{Skinner}, P.~{Jean}, F.~{Sanchez}, P.~{Caraveo}, B.~{Teegarden}, P.~{von
  Ballmoos}, L.~{Bouchet}, P.~{Paul}, J.~{Matteson}, S.~{Boggs}, C.~{Wunderer},
  P.~{Leleux}, G.~{Weidenspointner}, P.~{Durouchoux}, R.~{Diehl}, A.~{Strong},
  M.~{Cass{\'e}}, M.~A. {Clair} and Y.~{Andr{\'e}}, \emph{{SPI: The
  spectrometer aboard INTEGRAL}},
  \href{https://doi.org/10.1051/0004-6361:20031482}{\emph{A\&A} {\bfseries 411}
  (2003) L63}.

\bibitem{BeechertCOSICalib}
J.~{Beechert}, H.~{Lazar}, S.~E. {Boggs}, T.~J. {Brandt}, Y.-C. {Chang}, C.-Y.
  {Chu}, H.~{Gulick}, C.~{Kierans}, A.~{Lowell}, N.~{Pellegrini}, J.~M.
  {Roberts}, T.~{Siegert}, C.~{Sleator}, J.~A. {Tomsick} and A.~{Zoglauer},
  \emph{{Calibrations of the Compton Spectrometer and Imager}},
  \href{https://doi.org/10.1016/j.nima.2022.166510}{\emph{Nuclear Instruments
  and Methods in Physics Research A} {\bfseries 1031} (2022) 166510}
  [\href{https://arxiv.org/abs/2203.00695}{{\ttfamily 2203.00695}}].

\bibitem{Boehm:2002yz}
C.~Boehm, T.~A. Ensslin and J.~Silk, \emph{{Can Annihilating dark matter be
  lighter than a few GeVs?}},
  \href{https://doi.org/10.1088/0954-3899/30/3/004}{\emph{J. Phys. G}
  {\bfseries 30} (2004) 279}
  [\href{https://arxiv.org/abs/astro-ph/0208458}{{\ttfamily
  astro-ph/0208458}}].

\bibitem{Beacom:2004pe}
J.~F. Beacom, N.~F. Bell and G.~Bertone, \emph{{Gamma-ray constraint on
  Galactic positron production by MeV dark matter}},
  \href{https://doi.org/10.1103/PhysRevLett.94.171301}{\emph{Phys. Rev. Lett.}
  {\bfseries 94} (2005) 171301}
  [\href{https://arxiv.org/abs/astro-ph/0409403}{{\ttfamily
  astro-ph/0409403}}].

\bibitem{Finkbeiner:2007kk}
D.~P. Finkbeiner and N.~Weiner, \emph{{Exciting Dark Matter and the
  INTEGRAL/SPI 511 keV signal}},
  \href{https://doi.org/10.1103/PhysRevD.76.083519}{\emph{Phys. Rev. D}
  {\bfseries 76} (2007) 083519}
  [\href{https://arxiv.org/abs/astro-ph/0702587}{{\ttfamily
  astro-ph/0702587}}].

\bibitem{Essig:2009jx}
R.~Essig, N.~Sehgal and L.~E. Strigari, \emph{{Bounds on Cross-sections and
  Lifetimes for Dark Matter Annihilation and Decay into Charged Leptons from
  Gamma-ray Observations of Dwarf Galaxies}},
  \href{https://doi.org/10.1103/PhysRevD.80.023506}{\emph{Phys. Rev. D}
  {\bfseries 80} (2009) 023506}
  [\href{https://arxiv.org/abs/0902.4750}{{\ttfamily 0902.4750}}].

\bibitem{Essig:2013goa}
R.~Essig, E.~Kuflik, S.~D. McDermott, T.~Volansky and K.~M. Zurek,
  \emph{{Constraining Light Dark Matter with Diffuse X-Ray and Gamma-Ray
  Observations}}, \href{https://doi.org/10.1007/JHEP11(2013)193}{\emph{JHEP}
  {\bfseries 11} (2013) 193} [\href{https://arxiv.org/abs/1309.4091}{{\ttfamily
  1309.4091}}].

\bibitem{Boddy:2015efa}
K.~K. Boddy and J.~Kumar, \emph{{Indirect Detection of Dark Matter Using
  MeV-Range Gamma-Ray Telescopes}},
  \href{https://doi.org/10.1103/PhysRevD.92.023533}{\emph{Phys. Rev. D}
  {\bfseries 92} (2015) 023533}
  [\href{https://arxiv.org/abs/1504.04024}{{\ttfamily 1504.04024}}].

\bibitem{Carr:2020xqk}
B.~Carr and F.~Kuhnel, \emph{{Primordial Black Holes as Dark Matter: Recent
  Developments}},
  \href{https://doi.org/10.1146/annurev-nucl-050520-125911}{\emph{Ann. Rev.
  Nucl. Part. Sci.} {\bfseries 70} (2020) 355}
  [\href{https://arxiv.org/abs/2006.02838}{{\ttfamily 2006.02838}}].

\bibitem{Green:2020jor}
A.~M. Green and B.~J. Kavanagh, \emph{{Primordial Black Holes as a dark matter
  candidate}}, \href{https://doi.org/10.1088/1361-6471/abc534}{\emph{J. Phys.
  G} {\bfseries 48} (2021) 043001}
  [\href{https://arxiv.org/abs/2007.10722}{{\ttfamily 2007.10722}}].

\bibitem{Sasaki:2018dmp}
M.~Sasaki, T.~Suyama, T.~Tanaka and S.~Yokoyama, \emph{{Primordial black
  holes\textemdash{}perspectives in gravitational wave astronomy}},
  \href{https://doi.org/10.1088/1361-6382/aaa7b4}{\emph{Class. Quant. Grav.}
  {\bfseries 35} (2018) 063001}
  [\href{https://arxiv.org/abs/1801.05235}{{\ttfamily 1801.05235}}].

\bibitem{Villanueva-Domingo:2021spv}
P.~Villanueva-Domingo, O.~Mena and S.~Palomares-Ruiz, \emph{{A brief review on
  primordial black holes as dark matter}},
  \href{https://doi.org/10.3389/fspas.2021.681084}{\emph{Front. Astron. Space
  Sci.} {\bfseries 8} (2021) 87}
  [\href{https://arxiv.org/abs/2103.12087}{{\ttfamily 2103.12087}}].

\bibitem{Ray:2021mxu}
A.~Ray, R.~Laha, J.~B. Mu\~noz and R.~Caputo, \emph{{Near future MeV telescopes
  can discover asteroid-mass primordial black hole dark matter}},
  \href{https://doi.org/10.1103/PhysRevD.104.023516}{\emph{Phys. Rev. D}
  {\bfseries 104} (2021) 023516}
  [\href{https://arxiv.org/abs/2102.06714}{{\ttfamily 2102.06714}}].

\bibitem{Knodlseder:2005yq}
J.~Knodlseder et~al., \emph{{The All-sky distribution of 511 keV
  electron-positron annihilation emission}},
  \href{https://doi.org/10.1051/0004-6361:20042063}{\emph{Astron. Astrophys.}
  {\bfseries 441} (2005) 513}
  [\href{https://arxiv.org/abs/astro-ph/0506026}{{\ttfamily
  astro-ph/0506026}}].

\bibitem{Laha:2020ivk}
R.~Laha, J.~B. Mu\~noz and T.~R. Slatyer, \emph{{INTEGRAL constraints on
  primordial black holes and particle dark matter}},
  \href{https://doi.org/10.1103/PhysRevD.101.123514}{\emph{Phys. Rev. D}
  {\bfseries 101} (2020) 123514}
  [\href{https://arxiv.org/abs/2004.00627}{{\ttfamily 2004.00627}}].

\bibitem{Calore:2022pks}
F.~Calore, A.~Dekker, P.~D. Serpico and T.~Siegert, \emph{{Constraints on light
  decaying dark matter candidates from 16 years of INTEGRAL/SPI observations}},
   \href{https://arxiv.org/abs/2209.06299}{{\ttfamily 2209.06299}}.

\bibitem{Tomsick:2021wed}
{\scshape COSI} Collaboration, J.~A. Tomsick, \emph{{The Compton Spectrometer
  and Imager Project for MeV Astronomy}},
  \href{https://doi.org/10.22323/1.395.0652}{\emph{PoS} {\bfseries ICRC2021}
  (2021) 652} [\href{https://arxiv.org/abs/2109.10403}{{\ttfamily
  2109.10403}}].

\bibitem{Regis:2021glv}
M.~Regis et~al., \emph{{The EMU view of the Large Magellanic Cloud: troubles
  for sub-TeV WIMPs}},
  \href{https://doi.org/10.1088/1475-7516/2021/11/046}{\emph{JCAP} {\bfseries
  11} (2021) 046} [\href{https://arxiv.org/abs/2106.08025}{{\ttfamily
  2106.08025}}].

\bibitem{Karukes:2019jxv}
E.~V. Karukes, M.~Benito, F.~Iocco, R.~Trotta and A.~Geringer-Sameth,
  \emph{{Bayesian reconstruction of the Milky Way dark matter distribution}},
  \href{https://doi.org/10.1088/1475-7516/2019/09/046}{\emph{JCAP} {\bfseries
  09} (2019) 046} [\href{https://arxiv.org/abs/1901.02463}{{\ttfamily
  1901.02463}}].

\bibitem{Cirelli:2010xx}
M.~Cirelli, G.~Corcella, A.~Hektor, G.~Hutsi, M.~Kadastik, P.~Panci, M.~Raidal,
  F.~Sala and A.~Strumia, \emph{{PPPC 4 DM ID: A Poor Particle Physicist
  Cookbook for Dark Matter Indirect Detection}},
  \href{https://doi.org/10.1088/1475-7516/2012/10/E01}{\emph{JCAP} {\bfseries
  03} (2011) 051} [\href{https://arxiv.org/abs/1012.4515}{{\ttfamily
  1012.4515}}]. [Erratum: JCAP 10, E01 (2012)].

\bibitem{Bonnivard:2015xpq}
V.~Bonnivard et~al., \emph{{Dark matter annihilation and decay in dwarf
  spheroidal galaxies: The classical and ultrafaint dSphs}},
  \href{https://doi.org/10.1093/mnras/stv1601}{\emph{Mon. Not. Roy. Astron.
  Soc.} {\bfseries 453} (2015) 849}
  [\href{https://arxiv.org/abs/1504.02048}{{\ttfamily 1504.02048}}].

\bibitem{Negro:2021urm}
M.~Negro, H.~Fleischhack, A.~Zoglauer, S.~Digel and M.~Ajello, \emph{{Unveiling
  the Origin of the Fermi Bubbles with MeV Photon Telescopes}},
  \href{https://doi.org/10.3847/1538-4357/ac5326}{\emph{Astrophys. J.}
  {\bfseries 927} (2022) 225}
  [\href{https://arxiv.org/abs/2111.10362}{{\ttfamily 2111.10362}}].

\bibitem{Berteaud:2022tws}
J.~Berteaud, F.~Calore, J.~Iguaz, P.~D. Serpico and T.~Siegert, \emph{{Strong
  constraints on primordial black hole dark matter from 16~years of
  INTEGRAL/SPI observations}},
  \href{https://doi.org/10.1103/PhysRevD.106.023030}{\emph{Phys. Rev. D}
  {\bfseries 106} (2022) 023030}
  [\href{https://arxiv.org/abs/2202.07483}{{\ttfamily 2202.07483}}].

\bibitem{Langhoff:2022bij}
K.~Langhoff, N.~J. Outmezguine and N.~L. Rodd, \emph{{The Irreducible Axion
  Background}},  \href{https://arxiv.org/abs/2209.06216}{{\ttfamily
  2209.06216}}.

\bibitem{Foster:2022ajl}
J.~W. Foster, S.~Kumar, B.~R. Safdi and Y.~Soreq, \emph{{Dark Grand Unification
  in the Axiverse: Decaying Axion Dark Matter and Spontaneous Baryogenesis}},
  \href{https://arxiv.org/abs/2208.10504}{{\ttfamily 2208.10504}}.

\bibitem{Aramaki:2022zpw}
T.~Aramaki et~al., \emph{{Snowmass2021 Cosmic Frontier: The landscape of
  cosmic-ray and high-energy photon probes of particle dark matter}},
  \href{https://arxiv.org/abs/2203.06894}{{\ttfamily 2203.06894}}.

\bibitem{Hawking:1974rv}
S.~W. Hawking, \emph{{Black hole explosions}},
  \href{https://doi.org/10.1038/248030a0}{\emph{Nature} {\bfseries 248} (1974)
  30}.

\bibitem{Arbey:2019mbc}
A.~Arbey and J.~Auffinger, \emph{{BlackHawk: A public code for calculating the
  Hawking evaporation spectra of any black hole distribution}},
  \href{https://doi.org/10.1140/epjc/s10052-019-7161-1}{\emph{Eur. Phys. J. C}
  {\bfseries 79} (2019) 693}
  [\href{https://arxiv.org/abs/1905.04268}{{\ttfamily 1905.04268}}].

\bibitem{Coogan:2020tuf}
A.~Coogan, L.~Morrison and S.~Profumo, \emph{{Direct Detection of Hawking
  Radiation from Asteroid-Mass Primordial Black Holes}},
  \href{https://doi.org/10.1103/PhysRevLett.126.171101}{\emph{Phys. Rev. Lett.}
  {\bfseries 126} (2021) 171101}
  [\href{https://arxiv.org/abs/2010.04797}{{\ttfamily 2010.04797}}].

\bibitem{Carr:2017jsz}
B.~Carr, M.~Raidal, T.~Tenkanen, V.~Vaskonen and H.~Veerm\"ae,
  \emph{{Primordial black hole constraints for extended mass functions}},
  \href{https://doi.org/10.1103/PhysRevD.96.023514}{\emph{Phys. Rev. D}
  {\bfseries 96} (2017) 023514}
  [\href{https://arxiv.org/abs/1705.05567}{{\ttfamily 1705.05567}}].

\bibitem{Cirelli:2020bpc}
M.~Cirelli, N.~Fornengo, B.~J. Kavanagh and E.~Pinetti, \emph{{Integral X-ray
  constraints on sub-GeV Dark Matter}},
  \href{https://doi.org/10.1103/PhysRevD.103.063022}{\emph{Phys. Rev. D}
  {\bfseries 103} (2021) 063022}
  [\href{https://arxiv.org/abs/2007.11493}{{\ttfamily 2007.11493}}].

\bibitem{Bystritskiy:2005ib}
Y.~M. Bystritskiy, E.~A. Kuraev, G.~V. Fedotovich and F.~V. Ignatov, \emph{{The
  Cross sections of the muons and charged pions pairs production at
  electron-positron annihilation near the threshold}},
  \href{https://doi.org/10.1103/PhysRevD.72.114019}{\emph{Phys. Rev. D}
  {\bfseries 72} (2005) 114019}
  [\href{https://arxiv.org/abs/hep-ph/0505236}{{\ttfamily hep-ph/0505236}}].

\bibitem{Coogan:2021rez}
A.~Coogan, A.~Moiseev, L.~Morrison and S.~Profumo, \emph{{Hunting for Dark
  Matter and New Physics with (a) GECCO}},
  \href{https://arxiv.org/abs/2101.10370}{{\ttfamily 2101.10370}}.

\bibitem{Slatyer:2016qyl}
T.~R. Slatyer and C.-L. Wu, \emph{{General Constraints on Dark Matter Decay
  from the Cosmic Microwave Background}},
  \href{https://doi.org/10.1103/PhysRevD.95.023010}{\emph{Phys. Rev. D}
  {\bfseries 95} (2017) 023010}
  [\href{https://arxiv.org/abs/1610.06933}{{\ttfamily 1610.06933}}].

\bibitem{1972ApJ...172L...1J}
I.~{Johnson}, W.~N., J.~{Harnden}, F.~R. and R.~C. {Haymes}, \emph{{The
  Spectrum of Low-Energy Gamma Radiation from the Galactic-Center Region.}},
  \href{https://doi.org/10.1086/180878}{\emph{ApJL} {\bfseries 172} (1972) L1}.

\bibitem{1978ApJ...225L..11L}
M.~{Leventhal}, C.~J. {MacCallum} and P.~D. {Stang}, \emph{{Detection of 511
  keV positron annihilation radiation from the galactic center direction.}},
  \href{https://doi.org/10.1086/182782}{\emph{ApJL} {\bfseries 225} (1978)
  L11}.

\bibitem{Prantzos:2010wi}
N.~Prantzos et~al., \emph{{The 511 keV emission from positron annihilation in
  the Galaxy}}, \href{https://doi.org/10.1103/RevModPhys.83.1001}{\emph{Rev.
  Mod. Phys.} {\bfseries 83} (2011) 1001}
  [\href{https://arxiv.org/abs/1009.4620}{{\ttfamily 1009.4620}}].

\bibitem{Laha:2019ssq}
R.~Laha, \emph{{Primordial Black Holes as a Dark Matter Candidate Are Severely
  Constrained by the Galactic Center 511 keV $\gamma$ -Ray Line}},
  \href{https://doi.org/10.1103/PhysRevLett.123.251101}{\emph{Phys. Rev. Lett.}
  {\bfseries 123} (2019) 251101}
  [\href{https://arxiv.org/abs/1906.09994}{{\ttfamily 1906.09994}}].

\bibitem{DeRocco:2019fjq}
W.~DeRocco and P.~W. Graham, \emph{{Constraining Primordial Black Hole
  Abundance with the Galactic 511 keV Line}},
  \href{https://doi.org/10.1103/PhysRevLett.123.251102}{\emph{Phys. Rev. Lett.}
  {\bfseries 123} (2019) 251102}
  [\href{https://arxiv.org/abs/1906.07740}{{\ttfamily 1906.07740}}].

\bibitem{Fuller:2017uyd}
G.~M. Fuller, A.~Kusenko and V.~Takhistov, \emph{{Primordial Black Holes and
  $r$-Process Nucleosynthesis}},
  \href{https://doi.org/10.1103/PhysRevLett.119.061101}{\emph{Phys. Rev. Lett.}
  {\bfseries 119} (2017) 061101}
  [\href{https://arxiv.org/abs/1704.01129}{{\ttfamily 1704.01129}}].

\bibitem{Beacom:2005qv}
J.~F. Beacom and H.~Yuksel, \emph{{Stringent constraint on galactic positron
  production}},
  \href{https://doi.org/10.1103/PhysRevLett.97.071102}{\emph{Phys. Rev. Lett.}
  {\bfseries 97} (2006) 071102}
  [\href{https://arxiv.org/abs/astro-ph/0512411}{{\ttfamily
  astro-ph/0512411}}].

\bibitem{Vincent:2012an}
A.~C. Vincent, P.~Martin and J.~M. Cline, \emph{{Interacting dark matter
  contribution to the Galactic 511 keV gamma ray emission: constraining the
  morphology with INTEGRAL/SPI observations}},
  \href{https://doi.org/10.1088/1475-7516/2012/04/022}{\emph{JCAP} {\bfseries
  04} (2012) 022} [\href{https://arxiv.org/abs/1201.0997}{{\ttfamily
  1201.0997}}].

\bibitem{Ascasibar:2005rw}
Y.~Ascasibar, P.~Jean, C.~Boehm and J.~Knoedlseder, \emph{{Constraints on dark
  matter and the shape of the Milky Way dark halo from the 511-keV line}},
  \href{https://doi.org/10.1111/j.1365-2966.2006.10226.x}{\emph{Mon. Not. Roy.
  Astron. Soc.} {\bfseries 368} (2006) 1695}
  [\href{https://arxiv.org/abs/astro-ph/0507142}{{\ttfamily
  astro-ph/0507142}}].

\end{thebibliography}\endgroup

\end{document}